\documentclass[10pt,conference]{IEEEtran}
\IEEEoverridecommandlockouts
\usepackage{algorithm}
\usepackage{algorithmicx}
\usepackage{algpseudocode}
\usepackage{cite}
\usepackage[utf8x]{inputenc} 
\usepackage{amsmath}
\usepackage{amssymb,amsmath}
\usepackage{graphicx}
\usepackage{epsfig}
\usepackage{grffile}
\usepackage{balance,color}
\usepackage{caption}
\usepackage{subcaption}
\usepackage{xcolor, soul}
\sethlcolor{cyan}
\providecommand{\keywords}[1]{{\textit{Index Terms}}}
\abovecaptionskip 
\belowcaptionskip 

\begin{document}
\title{ Edge-Enabled UAV Swarm Deployment for Rapid Post-Disaster Search and Rescue 
\vspace{-0.4em}  
} 
\author{ Alaa Awad Abdellatif, Helder Fontes, Andre Coelho, Luis M. Pessoa, Rui Campos   
\\
\begin{tabular}{c}
 INESC TEC and Faculdade de Engenharia, Universidade do Porto, Portugal \\  
		\thanks { This work was carried out within the scope of the HURRICANE project, which has received funding from the European Union's Horizon Europe research and innovation program under Grant Agreement 101168017.   }  
\end{tabular}
\vspace{-1.9em} 

}
\maketitle

\begin{abstract}
This paper presents an optimized Joint Radar-Communication (JRC) system utilizing multiple Unmanned Aerial Vehicles (UAVs) to simultaneously achieve sensing and communication objectives. By leveraging UAVs equipped with dual radar and communication capabilities, the proposed framework aims to maximize radar sensing performance across all UAVs in challenging environments. The proposed approach focuses on formulating and solving a UAV positioning and power allocation problem to optimize multi-UAV sensing and communications performance over multiple targets within designated zones.  Due to the NP-hard and combinatorial nature of the problem, we propose a Distributed JRC-based (DJRC) solution.  This solution employs an efficient reward for potential actions and consistently selects the best action that maximizes the reward while ensuring both communications and sensing performance. 
Simulation results demonstrate significant performance improvements of the proposed solution over state-of-the-art radar- or communication-centric trajectory planning methods, with polynomial complexity dependent on the number of UAVs and linear dependence on the iteration count.

\end{abstract}

\section{Introduction\label{sec:Introduction}} 

In 6G mobile communication systems, advanced technologies will address spectral congestion by enabling multiple applications to coexist within the same frequency bands. Joint Radar-Communication (JRC) systems are an example of this approach. The adoption of higher frequency bands, wider bandwidths, and massive antenna arrays will enable high-accuracy, high-resolution sensing, seamlessly integrating wireless sensing and communications into a single system for mutual benefit. 
This evolution drives the concept of ``network as a sensor'', leveraging communication networks for sensing tasks. Radio signals transmitted, received, and reflected by network elements can be utilized to interpret the physical environment, supporting services like precise localization, gesture recognition, activity detection, passive object tracking, and environmental reconstruction \cite{bayesteh2022integrated}.

We argue that leveraging Unmanned Aerial Vehicles (UAVs) equipped with dual-function radar and communications capabilities can further complement this vision by offering reliable connectivity and advanced sensing. These features make UAVs particularly suited for applications such as surveillance, disaster response, and complex environmental monitoring  \cite{9712460, abdellatif2024pdsr}.   
On the one hand, integrating radar and communications functionalities in multi-UAV JRC systems optimizes spectrum usage, reduces interference, and boosts overall performance. UAVs in JRC systems can adaptively plan trajectories and optimize power control, ensuring efficient and flexible networks for a wide range of applications. On the other hand, UAVs equipped with high-frequency radar outperform cameras in specific scenarios due to their unique capabilities. Radar ensures all-weather functionality, operating effectively in fog, rain, snow, and darkness. Unlike cameras, it directly measures object velocity, providing precise data for applications like traffic monitoring and autonomous vehicles. Radar can also penetrate obstacles like fog and light rain, detecting obscured objects, and its immunity to lighting variations ensures consistent performance regardless of time or glare. Additionally, its extended detection range makes it well-suited for early warning and surveillance tasks. 

Despite the potential and advancements of multi-UAV JRC systems, real-world deployment is complicated by factors such as limited fleet size, restricted coverage area per UAV, and challenging environmental conditions \cite{9685143}.  These limitations make effective UAV deployment strategy critical for maximizing radar sensory and communications performance. 
Existing research has explored various aspects of JRC systems, including radar-communication spectrum sharing, trajectory planning and power control, adaptive beamforming and waveform diversity to transmit separate radar and communication streams, and distributed systems where multiple transmitters and receivers collaboratively perform radar and communications tasks \cite{liao2023robust, wang2019power, hong2021interference, ahmed2023system}. 
For example, trajectory planning studies focus on optimizing UAV movements to improve both radar coverage and communications quality by considering environmental factors and mission constraints \cite{hu2020aoi}. Meanwhile, power control research addresses the allocation of power between radar and communications subsystems to ensure both effective radar sensing and communications throughput \cite{9963915, ahmed2023optimized}. Together, these advancements aim to maximize the performance and efficiency of JRC systems, balancing radar detection capabilities and communications requirements. 
However, while these approaches enhance JRC system performance, many rely on centralized solutions, which often lack flexibility, scalability, and fault tolerance. Furthermore, existing learning-based methods typically use discrete action spaces, leading to time-consuming strategy optimization, or assume fixed UAV altitudes, which constrains performance and limits optimality.  

This work addresses this gap by proposing a Distributed JRC-based (DJRC) solution designed to maximize radar detection quality for all UAVs through optimized UAVs' location and power, while guaranteeing communications performance.  
In particular, we employ a swarm of UAVs organized in a hierarchical two-tier architecture to enable scalable  sensing and communications operations. High-altitude UAVs work as Flying Base Stations (FBS), covering large areas, while low-altitude UAVs, equipped with dual-functional radar-communications systems, handle target-specific sensing and communications with the FBS. This two-tiered approach ensures broad and rapid coverage of designated areas, efficient resource allocation, and minimized redundancy and overlap. The backhaul communication protocols for the FBS, and the mechanical energy consumed for UAVs movement, are outside the scope of this article. 
Thus, our contributions can be summarized as follows: 
\begin{itemize}
    \item \textbf{Trajectory planning and power control:} We formulate an optimization problem for the proposed multi-tiered scenario to jointly optimize UAV locations in 3D space and power allocation between radar and communication functions. The objective is to maximize detection performance while ensuring constraints on radar parameter estimation accuracy and communications quality are met. 
    \item \textbf{Distributed optimization:} Given the NP-hard and combinatorial nature of the formulated problem, we propose an efficient distributed solution based on optimization decomposition and an optimality-driven reward mechanism. Our solution defines the UAVs' locations and power allocation while balancing detection accuracy and communications performance across multiple targets. 
    \item \textbf{Improved detection quality and reduced complexity:} Our approach demonstrates superior detection quality compared to traditional methods while significantly reducing computational complexity by eliminating the need for exhaustive searches. This efficiency makes it practical for implementation across various UAV setups.   
\end{itemize}

The following sections present the multi-UAV JRC system model (Section \ref{sec:Sec2}), the performance metrics and problem formulation (Section \ref{sec:system}), the proposed distributed JRC solution   (Section \ref{sec:Solutions}), the performance evaluation results (Section \ref{sec:Evaluation}), and the conclusions (Section \ref{sec:conclusion}).  


\section{Multi-UAV JRC System Model \label{sec:Sec2}}

{ 
The JRC enabled multi-UAV target detection system, illustrated in Figure~\ref{fig:system_model}, is designed to perform efficient aerial surveys, enhance sensing operations, and forward collected data to a First Responder Center (FRC) for analysis and decision-making. This system integrates two primary components: a high-altitude UAV acting as a FBS, and a swarm of $M$ identical low-altitude UAVs equipped with dual radar and communications capabilities. 
These UAVs perform sensing operations for $N$ stationary targets over $T$ consecutive time slots, and transmit the sensed data to the FBS, utilizing separate signals for sensing and communication.  
The FBS, in turn, receives and aggregates data from all attached UAVs, and forwards them to the FRC. 
The latter serves as the central command for UAV operations, overseeing deployment, coordinating tasks, and monitoring UAV status. 

The considered multi-tiered architecture allows the UAVs to fly at low altitudes, while the FBS operates at a higher altitude. This configuration improves radar detection quality and resolution for the UAVs, while also enhancing communications between the UAVs and FBS by improving Line-of-Sight (LoS), minimizing the impact of obstacles, and expanding the coverage area. 
However, flying at lower altitudes reduces the coverage range of each individual UAV, requiring the deployment of a larger number of UAVs to effectively cover a wide area. Furthermore, operating the FBS at very high altitudes may result in signal intensity loss due to path loss, and environmental factors, which can degrade the performance. 
This highlights the need for careful design and optimization of the UAV assignment strategy to ensure efficient resource utilization. The placement and coordination of UAVs must be optimized to achieve a balance between maximizing detection quality and maintaining communication performance with the FBS. 

Once the optimal location for each UAV is determined, they autonomously navigate to these positions for sensing and communication tasks. It is assumed that autonomous navigation and obstacle avoidance are integrated into each UAV, allowing them to independently activate their navigation systems at the start of the mission and navigate along the safest, most optimal routes, minimizing travel time and enhancing overall mission effectiveness.  

\begin{figure}[t!]
	\centering
		\scalebox{1.7}{\includegraphics[width=0.33 \textwidth]{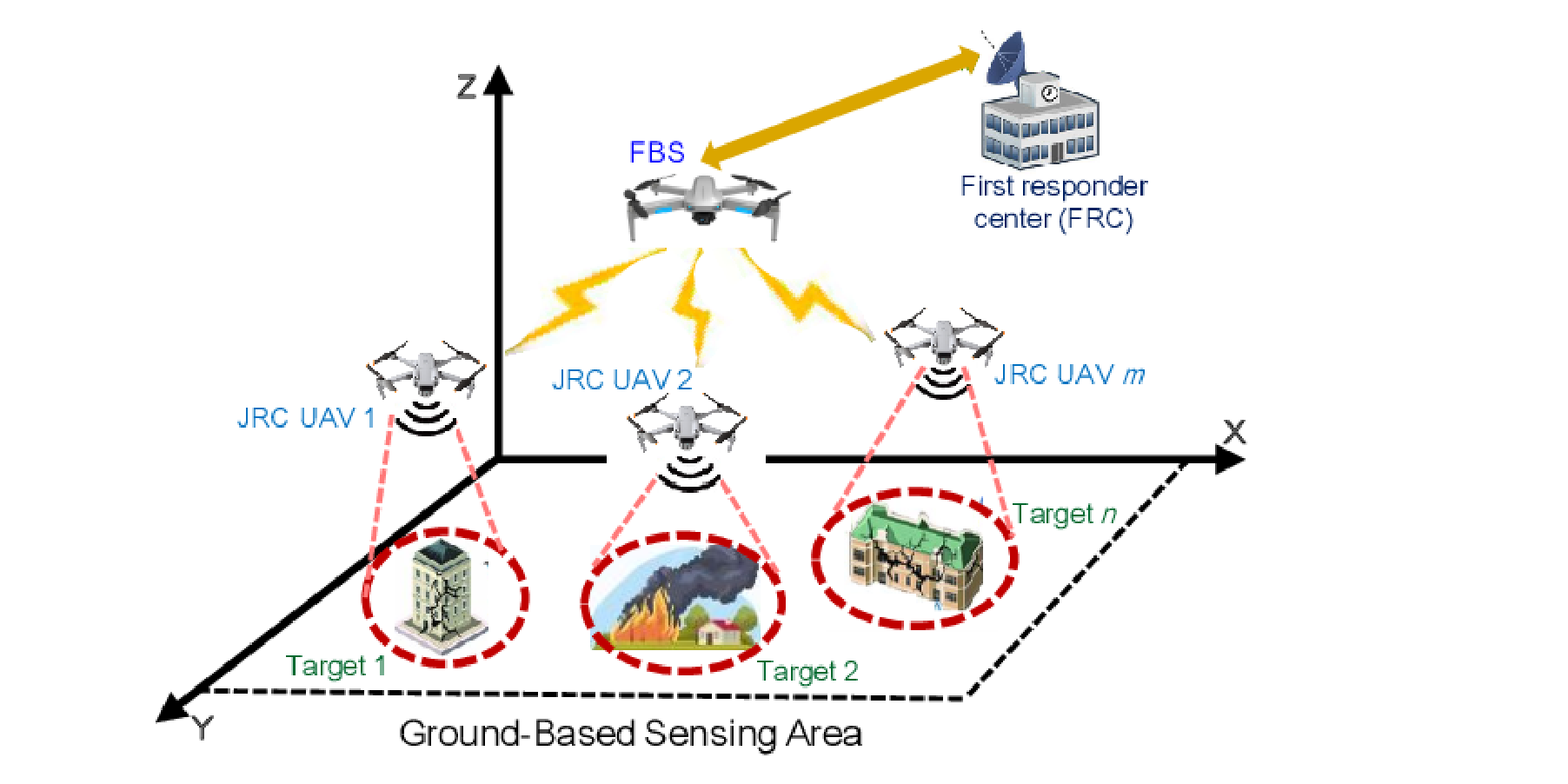}}
	\caption{The considered multi-UAV JRC system model. }
	\label{fig:system_model}
\end{figure}

In the considered system, each UAV employs separate hardware for sensing and communication, using distinct signals for each function. 
To avoid interference between these functions, each UAV is assigned a dedicated radio channel for radar sensing and a separate radio channel for communication \cite{8943325}. 
The coordinates of the $m$-th UAV at time slot $t$ are represented as $(x_m(t), y_m(t), H_m(t))$, determined via GPS. The coordinates of the FBS are denoted as $(x_h(t), y_h(t), H_h(t))$, where $H_h(t) > H_m(t)$. The coordinates of the $n$-th target are given by $(x_n, y_n, 0)$. Without loss of generality, the positions of all targets are assumed to remain fixed. 
The UAV-target assignments are pre-defined by the FRC, with each UAV designated to detect a single target at time slot $t$. Moreover, it is assumed that both radar and communication channels remain relatively constant within a time slot, allowing each UAV to use pilot signals for estimating its channel state information (CSI) \cite{9963915}. 
Finally, the UAVs are assumed to be fully charged, ensuring sufficient energy to reach their assigned locations and complete sensing and communication tasks. The mechanical energy consumed for UAV movement is beyond the scope of this work. Additionally, the power split between radar and communication pertains only to transmitted power, not the total power consumption of these systems. 
\section{Performance Metrics and Problem Formulation  \label{sec:system}}

In this section, we define the key performance metrics, including the radar sensing and communication models, and formulate the multi-UAV JRC optimization problem.

\subsection{ Radar Sensing Model \label{sec:sensing}}

We adopt the Radio Frequency (RF) radar model for sensing due to its flexibility and robustness in diverse operational environments. RF radar ensures reliable target detection and tracking in low visibility conditions like fog, rain, and darkness, where optical sensors may fail. It offers long-range sensing, precise velocity estimation, and seamless integration with communication systems, making it ideal for UAV-based JRC applications.  

To measure the radar detection quality, considering monostatic radar\footnote{In monostatic radar, the range from the target to the transmitter and receiver is identical.}, we consider the signal-to-noise ratio (SNR) between the $m$-th UAV and $n$-th target at time slot $t$, which is given by \cite{barton2013radar}:
\begin{equation}
\eta_{mn}(t) = \frac{p_m^r(t) g_m^T g_m^R \lambda^2 \sigma_n}{(4 \pi)^3 \Gamma B_m^r d_{mn}(t)^4}, 
\label{eq:SNR}
\end{equation}
where $p_m^r$ and $B_m^r$ is the allocated power and bandwidth (BW) for the radar of the $m$-th UAV, respectively, $g_m^T$ and $g_m^R$ refers to the transmitting and receiving antenna gains, respectively, $\lambda = C/f_c$ is the operating wavelength, and $\sigma_n$ is the Radar Cross Section (RCS) of the $n$-target. For high-frequency radars, $B_m^r << f_c$, where $f_c$ represents the operating frequency. ensuring effective signal modulation and detection.  
$\Gamma = k T_0 F l$, where $k$ is the Boltzmann constant, $T_0$ is the effective noise temperature in terms of thermodynamic temperature, $F$ and $l$ are the radar noise figure and probing loss, respectively \cite{9963915}.   
In (\ref{eq:SNR}), $d_{mn}$ is the the distance between the $m$-th UAV  
and the $n$-target on the ground, and it is calculated using the 3D Euclidean distance formula, as follows:
\begin{equation} 
d_{mn} = \sqrt{(x_m - x_n)^2 + (y_m - y_n)^2 + (H_m)^2}. 
\label{eq:dmn}
\end{equation}
To ensure the detection quality of the $m$-th UAV for the $n$-th target at time slot $t$, the obtained radar SNR should satisfy the following constraint:  
\begin{equation}  
\eta_{mn}(t) \geq \eta_{\text{min}},
\end{equation}
where $\eta_{\text{min}}$ represents the minimum SNR required for a UAV to detect a target in time slot $t$.  
Accordingly, the maximum detectable range of UAV $m$ to detect target $n$ in time slot $t$, known as the radar range, is defined as:      
\begin{equation}
R_{mn}(t) = \bigg( \frac{p_m^r(t) g_m^T g_m^R \lambda^2 \sigma_n}{(4 \pi)^3 \Gamma B_m^r  \eta_{\text{min}} } \bigg)^{1/4}. 
\label{eq:Range}
\end{equation}
This sets a constraint on the $m$-th UAV to guarantee the detection quality, i.e.,  
$ d_{mn}(t) \leq R_{mn}(t) $. 

\subsection{ Communication Model \label{sec:Communication}} 

While sensing the targets, each UAV needs to simultaneously transmit sensed data to the FBS. Following the communication model in \cite{9963915, qiu2020multiple, mozaffari2017mobile}, the average channel power gain between the $m$-th UAV and the FBS is calculated by:
\begin{equation}
h_{mh}(t) = K_0^{-1} d_{m0}(t)^{-2} \left[ \xi_m^{\text{LoS}}(t) \mu^{\text{LoS}} + \xi_m^{\text{NLoS}}(t) \mu^{\text{NLoS}} \right]^{-1}, 
\label{eq:COMMLoss}
\end{equation}
where $K_0 = (\frac{4 \pi f_c }{C})^2$, $\mu^{\text{LoS}}$ and $\mu^{\text{NLoS}}$ are the attenuation factors for LoS and Non-Line-of-Sight (NLoS) links, respectively. $\xi_m^{\text{LoS}}(t)$ and $\xi_m^{\text{NLoS}}(t)$ are the LoS and NLoS probabilities between the $m$-th UAV and the FBS, respectively.  
The distance between the $m$-th UAV and the FBS at time slot $t$ is denoted as $d_{mh}(t)$. 
Based on the allocated transmit power for communication $p_m^c(t)$ and channel gain, the Signal-to-Noise-Interference-Ratio (SINR) between the $m$-th UAV and the FBS is defined as:
\begin{equation}
\Psi_{mh}(t) = \frac{p_m^c(t) \cdot g_m^T \cdot g_h^R  \cdot h_{mh}(t)} {\sum_{u \neq m} p_u^c(t) \cdot g_u^T \cdot g_h^R \cdot h_{uh}(t) + B_m^c \cdot \delta_0 },
\label{eq:COMM_SNR}
\end{equation}
where $B_m^c$ is the allocated BW for communication and $\delta_0$ is the thermal noise.   
The data rate of the $m$-th UAV on the communication link is calculated using Shannon's capacity formula:
\begin{equation}
r_m(t) = B_m^c \log_2(1 + \Psi_{mh}(t)). 
\label{eq:rate}
\end{equation}
Hence, to ensure the quality of data transmission between the 
UAV and the FBS, we set a threshold $R_{\text{min}}$ for the transmission data rate of each UAV, such that, 
\begin{equation}  
r_m(t) \geq R_{\text{min}}. 
\label{eq:rateConst}
\end{equation}
It is worth noting that successive interference cancellation techniques can be used to deal with the interference from the echo signal scattered from certain target to the FBS \cite{chiriyath2019novel}. 

\subsection{ Problem Formulation \label{sec:Problem}} 

In the considered scenario of a swarm of $\mathcal{M}$ JRC-enabled UAVs deployed to detect targets in a specified region and transmit their data to a FBS, our objective is to maximize the radar detection quality of all UAVs while ensuring their detectable range and communication data rates with the FBS. Achieving this requires optimizing, for each UAV, its location, to balance radar coverage and communication efficiency, and the power splitting factor to optimally allocate power between radar and communication functions.  
Let $p_t$ represent the total power of each UAV. The power allocated by the $m$-th UAV for sensing at time slot $t$ is given by  $ p_m^r(t) = (1 - \gamma_m(t)) \cdot p_t$, and the power allocated for communication is  $ p_m^c(t) =  \gamma_m(t) \cdot  p_t$,  where $\gamma_m(t) \in [0, 1]$ is the power split factor for the $m$-th UAV at time slot $t$. 
This scenario is mathematically formulated as an optimization problem to effectively address the intricate trade-offs involved and to devise an efficient strategy for the JRC-enabled UAV network, as detailed below: 
\begin{eqnarray}
\mbox{\bf P:} &&\max_{ \gamma, \mathbf{q}, \mathbf{L}^h} \ \  \sum_{i=1}^{M} w_i \cdot \eta_{mn}(t)  
	\label{eq:optimize_prob} \\
&& \mbox{subject to: }  \nonumber  \\ 
&& d_{mn}(t) \leq R_{mn}(t), \quad \forall m \in  \mathcal{M},  \label{eq:C1}  \\  
&&  r_m(t) \geq R_{\text{min}}, \quad \forall m \in  \mathcal{M},  \label{eq:C2}  \\   
&&  d_{mm'}(t) \geq d_g, \quad \forall m \in \mathcal{M} \land m \neq m',   \label{eq:C3}  \\  
&& X_{\text{min}} \leq x_m(t), x_h(t) \leq X_{\text{max}}, \quad \forall m \in  \mathcal{M},  \label{eq:C5}  \\   
&& Y_{\text{min}} \leq y_m(t), y_h(t) \leq Y_{\text{max}}, \quad \forall m \in  \mathcal{M},  \label{eq:C6}  \\   
&& 0 < H_m, H_h \leq H_{\text{max}},   \label{eq:C7}  \\ 
&& \gamma_m(t) \in [0,1], \quad \forall m \in  \mathcal{M},    \label{eq:C8}  
\end{eqnarray}
where $w_i$ is a weighting coefficient used to assign different priorities to the UAVs (e.g., based on the importance of the sensed targets), satisfying the condition $\sum_{i=1}^{M} w_i = 1$.   

The formulated optimization problem in \textbf{P} aims to determine the optimal locations of the UAVs $\mathbf{q}_m$ and the FBS $\mathbf{L}^h$ to maximize the detection quality across all targets. Here, $\mathbf{q}_m = (x_m, y_m, H_m) \mid m = \{1, 2, \dots, M\}$ represents the location of $m$-th UAV, and  $\mathbf{L}^h = (x_h, y_h, H_h)$ represents the location of the FBS, both defined in a 3D space.  
Thus, we aim to determine $\mathbf{p}_m$, $\mathbf{L}^h$, and the power splitting factor $\gamma$, while maximizing their detection quality and maintaining the required communication data rates with the FBS.  
The constraint in (\ref{eq:C1}) ensures the distance between the $m$-th UAV and the $n$-th target is within the radar range. 
The constraint in (\ref{eq:C2}) ensures the minimum transmission data rate for all UAVs is met.  
The constraint in (\ref{eq:C3}) maintains a safe distance $d_g$ between UAVs to prevent collisions, where the distance $d_{mm'}(t)$ between the \(m\)-th UAV and the \(m'\)-th UAV is calculated as,
\begin{equation}
  d_{mm'} = \sqrt{(x_m - x_{m'})^2 + (y_m - y_{m'} )^2 + (H_m - H_{m'} )^2 }.   
\end{equation}
The constraints in (\ref{eq:C5}), (\ref{eq:C6}), and (\ref{eq:C7}) guarantee that all UAV flights are restricted to a region defined by \([X_{\text{min}}, X_{\text{max}}] \times [Y_{\text{min}}, Y_{\text{max}}] \times H_{\text{max}} \). Finally, the constraint in (\ref{eq:C8}) ensures that the power splitting factor $\gamma_m$ for each UAV $m$ remains within the valid range of $[0,1]$.   

The formulated problem in \textbf{P} is NP-hard due to its non-convex and combinatorial nature \cite{woeginger2003exact}, as evidenced by constraints (\ref{eq:C2}) and (\ref{eq:C3}). Therefore, in the following section, we propose a Distributed Joint Radar and Communication (DJRC) optimization algorithm to address this challenge. 

\section{ Distributed JRC Solution 
  \label{sec:Solutions}} 

To solve the problem formulated in \textbf{P}, we first decompose the original problem into manageable sub-problems that can be solved efficiently. We then apply the Distributed Joint Radar-Communication (DJRC) algorithm to solve these sub-problems iteratively, continuing until convergence is achieved. The details of the problem decomposition and the proposed algorithm are provided in the following subsections.

\subsection{ Problem Decomposition }
 
In order to analytically solve the problem in \textbf{P}, we decompose it into two subproblems, each dependent on one or more decision variables and solvable independently \cite{abdellatif2018user}. The challenge arises from the coupling of optimization variables (i.e., $\mathbf{q}$, $\mathbf{L}^h$). To address this, we optimize the UAV  variables ($\gamma$ and $\mathbf{q}$) and the FBS variables ($\mathbf{L}^h$) separately. FBS variables are global, affecting the overall system, while UAV variables are local and can be optimized independently in a distributed manner. Thus, we decompose the problem into FBS and  UAV  subproblems as follows:
\begin{eqnarray}
\mbox{\bf SP1:} &&\max_{\mathbf{L}^h} \ \  \sum_{i=1}^{M} w_i \cdot \eta_{mn}(t)  
	\label{eq:optimize_prob_sp1} \\
&& \mbox{subject to: }  \nonumber  \\ 
&&  r_m(t) \geq R_{\text{min}}, \quad \forall m \in  \mathcal{M},  \label{eq:C12}  \\   
&& X_{\text{min}} \leq x_h(t) \leq X_{\text{max}}, \quad \forall m \in  \mathcal{M},  \label{eq:C15}  \\   
&& Y_{\text{min}} \leq y_h(t) \leq Y_{\text{max}}, \quad \forall m \in  \mathcal{M},  \label{eq:C16}  \\   
&& 0 <  H_h \leq H_{\text{max}},   \label{eq:C17} 
\end{eqnarray}
\begin{eqnarray}
\mbox{\bf SP2:} &&\max_{ \gamma, \mathbf{q} } \ \  \sum_{i=1}^{M} w_i \cdot \eta_{mn}(t)  
	\label{eq:optimize_prob_sp2} \\
&& \mbox{subject to: }  \nonumber  \\ 
&& (\ref{eq:C1}), (\ref{eq:C2}), (\ref{eq:C3}), (\ref{eq:C8}),  \label{eq:C11} \nonumber  \\  
&& X_{\text{min}} \leq x_m(t) \leq X_{\text{max}}, \quad \forall m \in  \mathcal{M},  \label{eq:C25}  \\   
&& Y_{\text{min}} \leq y_m(t) \leq Y_{\text{max}}, \quad \forall m \in  \mathcal{M},  \label{eq:C26}  \\   
&& 0 < H_m \leq H_{\text{max}},   \label{eq:C27}   
\end{eqnarray}
\subsection{FBS  Optimization}

By analyzing \textbf{SP1}, we observe that the optimization variable $\mathbf{L}^h$ does not influence the objective $\sum_{i=1}^{M} w_i \cdot \eta_{mn}(t)$. However, it does affect the constraint in (\ref{eq:C12}). The location of the FBS impacts the data rates ($r_m$) of the UAVs, but not their detection performance, which depends solely on their location and power. Therefore, the optimal location for the FBS is the one that maximizes the data rates of all UAVs. As a result, the problem in \textbf{SP1} is reformulated as follows: 
\begin{eqnarray}
\mbox{\bf {P2}:} &&\max_{\mathbf{L}^h} \ \  \sum_{i=1}^{M} r_m(t)
	\label{eq:optimize_prob2} \\
&& \mbox{subject to: }  \nonumber  \\ 
\nonumber &&  (\ref{eq:C15}), (\ref{eq:C16}), (\ref{eq:C17}),  \label{eq:C112}  
\end{eqnarray}

To solve \textbf{P2} and determine the optimal location for the FBS, $\mathbf{L}^h$, we employ a gradient-based method, as follows: 
\subsubsection{Initialization} 
To ensure balanced data rates between all UAVs and the FBS, the initial location of the FBS, denoted as $\mathbf{L}^h_{in} $, is determined as follows: 
\begin{equation}
 \mathbf{L}^h_{in}  = {\small \left( \sum_{n=1}^N \frac{x_n}{N},  \sum_{n=1}^N \frac{y_n}{N}, \max(H_m) + d_h \right)}, 
  \label{eq:FBSI} 
\end{equation}
where $d_h$ represents the minimum height difference between the maximum UAVs' height and the FBS to avoid collisions. This initial position represents the centroid of all target locations in the horizontal plane and at an altitude greater than the maximum UAV height by a safe margin.

\subsubsection{Gradient Calculation} Compute the gradients of the objective function in (\ref{eq:optimize_prob2}), i.e., 
$\nabla f(\mathbf{L^h})$, by taking the partial derivatives of $f(\mathbf{L^h})$ with respect to each coordinate of $\mathbf{L^h}$, such that:   
\begin{equation}
\nabla f(\mathbf{L^h}) = \left( \frac{\partial f}{\partial x}, \frac{\partial f}{\partial y}, \frac{\partial f}{\partial h} \right),
\end{equation}
where $ f(\mathbf{L^h}) = \sum_{i=1}^{M} r_m(t)$. 

\subsubsection{Update Rule} Update the FBS location $\mathbf{L^h}$  iteratively by using the gradient ascent rule, as follows:   
\begin{equation}
\mathbf{L^h}(t+1) = \mathbf{L^h}(t) + \alpha \nabla f(\mathbf{L^h}),
\end{equation}
where $\alpha > 0$ is the learning rate. 
It is important to ensure that the updated location $\mathbf{L^h}(t+1)$ remains within the feasible region defined by the constraints in equations (\ref{eq:C15}), (\ref{eq:C16}), (\ref{eq:C17}).

\subsubsection{Termination Criteria} 
Repeat steps 2-3 until convergence constraint is satisfied, i.e., 
\begin{equation}
\|\nabla f(\mathbf{L^h})\| < \epsilon,
\end{equation}
where $\epsilon$ is a predefined tolerance. 

We emphasize that our gradient-based solution  provides a practical and computationally efficient solution for FBS optimization, as it leverages the local gradient to guide the search for the optimal solution. This approach enables faster convergence compared to exhaustive search methods, which is particularly advantageous in large-scale 3D search spaces, such as the one in our case. 

\subsection{ UAV Optimization }

To maximize the objective function in \textbf{SP2}, each UAV should be as close as possible to its assigned target while utilizing the maximum available power for sensing. However, this must be done while satisfying the constraints in (\ref{eq:C1})-(\ref{eq:C8}). To achieve the optimal solution, we assume that each UAV $m$ initially positions itself at the closest possible point to its target, i.e., $x_m = x_n$, $y_m = y_n$, $H_m = d_g$, considering a safe distance $d_g$,  while allocating its entire power to sensing  ($\gamma_m=0$). If the initial configuration violates any constraints, the UAV can adjust its optimization variables by either  
(i) increasing $\gamma_m$ to balance sensing and communications power, or 
(ii) moving towards the FBS while maintaining the shortest possible distance to both the target and the FBS; this is achieved by moving toward the FBS along a spherical surface centered at the target and incrementally increasing the sphere’s radius by a small step $\Delta_r$. 

To mathematically formulate our solution, we define a reward function $\mathcal{R} (m)$ for each UAV. This reward function accounts for the increase in achieved rate $r_m$ and the decrease in $\eta_{mn}$, at each time step $t$, due to the taken action by a UAV. Hence, it is defined as: 
\begin{equation}
\mathcal{R} (m) =  \left( \frac{r_m^{t+1} - r_m^t}{r_m^t} \right)  - \left( \frac{\eta_{mn}^t - \eta_{mn}^{t+1}}{\eta_{mn}^t} \right) 
  \label{eq:reward}
\end{equation}
At each time step $t$, each UAV takes the action $A_m \in \{a_1, a_2\}$ that maximizes its reward $\mathcal{R} (m)$, such that $a_1$ refers to increasing communications power,  
 and  $a_2$ refers to  moving toward the FBS with step $\Delta_r$, which is formulated as follows: 
\[
A_m(t) = \arg\max_{A \in \{a_1, a_2\}} \mathcal{R}(m)
\]
with the following update rules: 
\begin{equation}
\begin{cases} 
\gamma_m(t+1) = \gamma_m(t) + \Delta_\gamma, & \text{if } A_m(t) = a_1 \\ 
\mathbf{q}_m(t+1) = \mathbf{q}_m(t) + \Delta_r \cdot \frac{\mathbf{L}^h - \mathbf{q}_m(t)}{\|\mathbf{L}^h - \mathbf{q}_m(t)\|}, & \text{if } A_m(t) = a_2
\end{cases}
  \label{eq:update}
\end{equation}

By using this formulation, the UAVs independently adjust their position or power splitting based on reward maximization. This iterative process ensures the UAVs select the optimal action, update their positions or power levels accordingly, and remain as close as possible to their target while moving toward the FBS. The process continues until all constraints in (\ref{eq:C1})-(\ref{eq:C8}) are satisfied.  
We argue that as long as the step sizes ($\Delta_\gamma$ and $\Delta_r$) are sufficiently small, the system gradually converges to the optimal configuration that maximizes the objective function in \textbf{SP2}, while ensuring all constraints are satisfied. 

\subsection{ Distributed Joint Radar and Communication (DJRC) }

In this section, we introduce DJRC, a fully distributed and iterative algorithm for optimal UAV positioning and power allocation. Building on the problem decomposition discussed in the previous subsections, DJRC is designed to efficiently solve the optimization problem in \textbf{P}, in practical scenarios where each UAV is aware only of its assigned target location. 

According to DJRC, each UAV is initially positioned as close as possible to its assigned target while utilizing its total power for sensing. To ensure balanced data rates among UAVs, the FBS is initially placed at the centroid of all target locations in the horizontal plane, with an altitude exceeding the maximum UAV height by a safe margin, as defined in (\ref{eq:FBSI}).   
Then, at each iteration, each UAV $m$ evaluates its reward function in (\ref{eq:reward}) to determine whether to increase its power split factor \( \gamma_m \) by \( \Delta_\gamma \) or move toward the FBS along the shortest path, as defined in (\ref{eq:update}). Once all UAVs have taken their respective actions, the FBS solves its optimization problem in \textbf{P2} to update its location, maximizing the overall data rates for all UAVs.  
This iterative process continues until convergence is achieved—i.e., all constraints in (\ref{eq:C1})-(\ref{eq:C8}) are satisfied—or until a predefined maximum number of iterations $T_{m}$ is reached. 

The main steps of the DJRC algorithm are presented in Algorithm \ref{alg:DJRC}. 
\begin{algorithm}[h]
\caption{Distributed Joint Radar-Communications (DJRC) Algorithm}
\label{alg:DJRC}
\begin{algorithmic}[1]
\State \textbf{Initialization:} $x_m = x_n$, $y_m = y_n$, $H_m = d_g$, and $\gamma_m=0$, $\forall m \in  \mathcal{M}$.  
\State Calculate the initial FBS position using the Equation  (\ref{eq:FBSI}). 
\For {$t=1$ to $T_{m}$ }
    \For{ $m=1$ to $M$ (in parallel)}
        \State Compute reward function $\mathcal{R} (m)$ using (\ref{eq:reward}).
        \State Select action $A_m$ that maximizes $\mathcal{R} (m)$:
        \If{$A_m = a_1$}
            \State Increase power split factor using (\ref{eq:update}).  
        \ElsIf{$A_m = a_2$}
            \State Move toward FBS using (\ref{eq:update}). 
        \EndIf
    \EndFor
\State \textbf{FBS Update:} Solve problem \textbf{P2} to update the FBS location $\mathbf{L}^h$. 
\If {All constraints in (\ref{eq:C1})-(\ref{eq:C8}) are satisfied} 
\State Break 
\EndIf 
\EndFor 
\State \textbf{Output:} Optimal UAV positions $\mathbf{q}_m$ and power split factors $\gamma_m$, and FBS location $\mathbf{L}^h$.
\end{algorithmic}
\end{algorithm}

The computational complexity of the DJRC algorithm can be estimated by the iterative updates of UAVs and the optimization of the FBS location. Each iteration consists of UAV updates, which involve computing the reward function and selecting an action, both requiring $O(1)$ operations per UAV, leading to a total of $O(M)$ complexity. The FBS update is performed by solving $\mathbf{P2}$ using a gradient-based method, which has a complexity of $O(T_F)$, where $T_F$ is the number of iterations required for convergence. Given that the outer loop runs for at most $T_m$ iterations, the total complexity of the DJRC algorithm is $O(T_m \cdot T_F \cdot M)$, indicating a polynomial dependence on the number of UAVs and a linear dependence on both iteration counts.  
We emphasize that this complexity applies if the DJRC algorithm is executed in a centralized manner. However, our decentralized approach distributes the computational burden across UAVs, resulting in a per-UAV complexity of \( O(T_m \cdot T_F) \).

\section{Performance Evaluation   \label{sec:Evaluation}}
{ 
In this section, we evaluate the effectiveness of our  proposed solution and analyze the impact of key parameters. First, we demonstrate the convergence of our approach. Then, we compare its performance against two representative state-of-the-art methods: Fixed Radar Optimized Communications (FROC) and Optimized Radar Fixed Communications (ORFC). 
The FROC method optimizes communications throughput for all UAVs while assuming the maximum radar range, without considering the joint nature of the system \cite{hashir2022rate}. Conversely, the ORFC method prioritizes radar performance by positioning UAVs to meet the minimum data rate constraints. Both methods assume fixed power allocation. They are evaluated under the same environmental conditions for a fair comparison with our proposed solution. 
    
Our performance evaluation is conducted under two scenarios: (i) varying the number of targets, and (ii) varying the total available power at each UAV. In each scenario, we evaluate and compare the total received SNR and the overall data rates across all UAVs for each method.
The main simulation parameters are summarized in Table \ref{tab:parameters}, and all simulations were carried out in MATLAB. 

\begin{table}[t!]
\centering
\caption{Simulation Parameters.}
\begin{tabular}{|l|l|l|l|}
\hline
\textbf{Parameter}          & \textbf{Value}                & \textbf{Parameter}          & \textbf{Value}                \\ \hline
$X_{\text{min}}$            &  0 m                              & $\mu_{\text{NLoS}}$                    &  2                       \\ \hline
$X_{\text{max}}$            & 1000 m                         & $g_m^T$                     & 20              \\ \hline
$d_g$                       & 40 m                           & $g_h^R$                     & 20               \\ \hline
$p_t$                  & $30$ W                        & $\sigma_n$                  & 1 m\(^2\)                      \\ \hline
   $B_m^c$           &    40 MHz                        & $f_c$                       & 5 GHz                          \\ \hline
 $B_m^r$            &  20 MHz                        & $C$                          & $3 \times 10^8$ m/s            \\ \hline
$Y_{\text{min}}$            & 0 m                            & $R_{\text{min}}$                   & 0.1  Mbit/s                        \\ \hline
$Y_{\text{max}}$            & 1000 m                         & $\delta_0$                  & $0.5 \times 10^{-10}$ W/Hz    \\ \hline
$H_{\text{max}}$            & 100 m                          & $k$                          & $1.38 \times 10^{-23}$ J/K    \\ \hline
$\xi_{m_{\text{LoS}}}$      & 0.95                           & $T_0$                        & 290 K                          \\ \hline
$\xi_{m_{\text{NLoS}}}$     & 0.5                            & $F$                          & 5 dB    \\ \hline
$\mu_{\text{LoS}}$          & 0.5                            & $l$                          & 0.8                            \\ \hline
\end{tabular}
\label{tab:parameters}
\end{table}

}
To illustrate the convergence behavior of our DJRC algorithm, we present Figures \ref{fig:Location3D} and \ref{fig:convergence}. In these figures, we consider a scenario where three UAVs monitor three targets, with an FBS supporting communications. 
Figure \ref{fig:Location3D} depicts the 3D trajectory of each UAV while executing DJRC. Initially, each UAV starts from the closest feasible position to its assigned target. It then gradually moves toward the FBS while adjusting its power split factor to achieve an optimal balance between sensing and communications. This iterative process continues until the UAVs reach their optimal locations, ensuring the best tradeoff between sensing and communications while satisfying all constraints.

 \begin{figure}[b!]
 \vspace{-20pt}
	\centering
		\scalebox{1.55}{\includegraphics[width=0.34 \textwidth, height=0.25\textwidth] {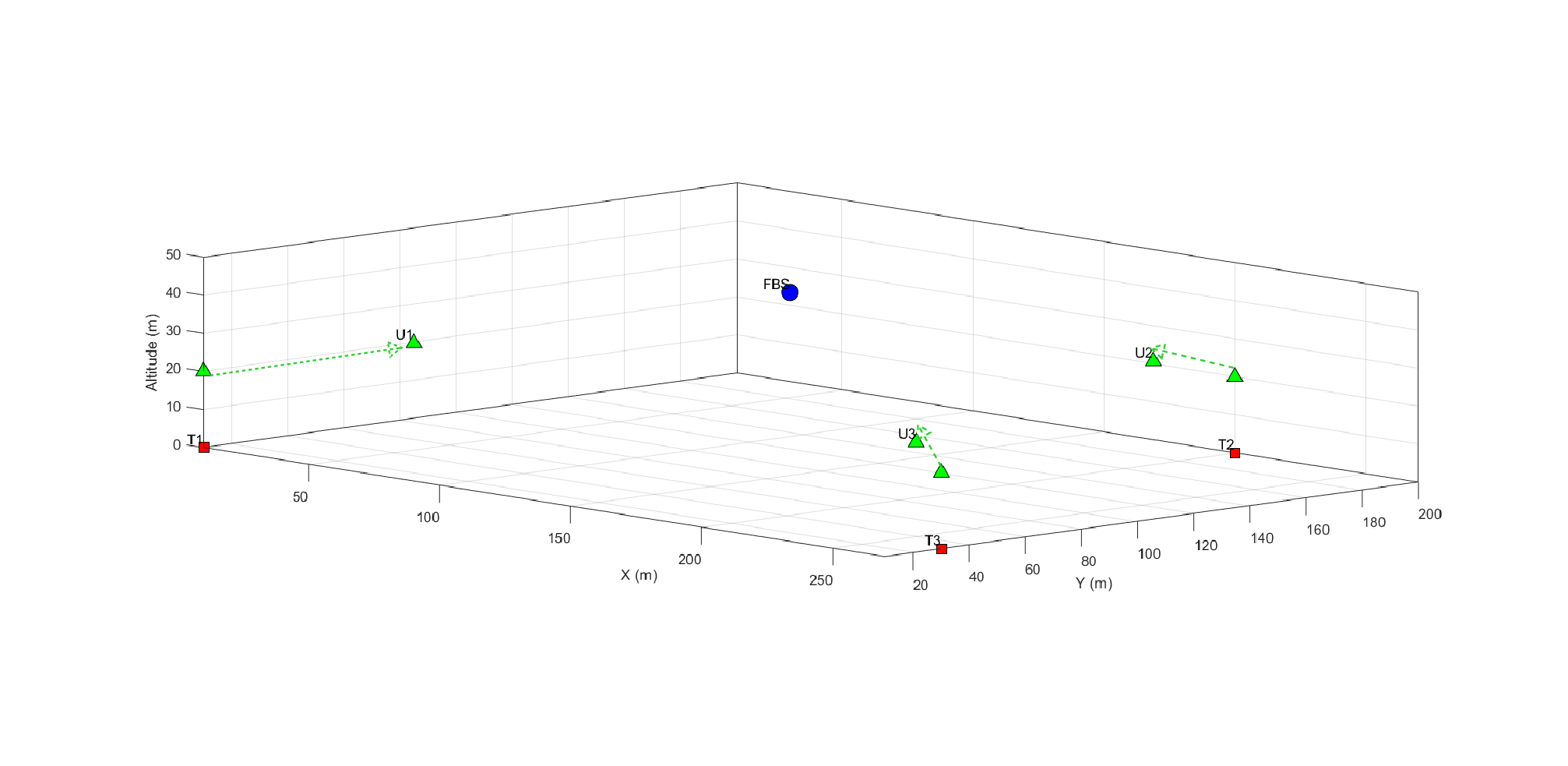} 
        }
	\caption{3D trajectory of UAVs during the execution of the DJRC algorithm.   } 
	\label{fig:Location3D}
\end{figure}

Figure \ref{fig:convergence} further illustrates the tradeoff between sensing and communications by depicting the evolution of radar performance—measured by the total received SNR across all UAVs ($\eta_t$)—and communications quality, represented by the sum of data rates achieved by all UAVs (${R}_t$), as the algorithm progresses toward convergence. Initially, since the UAVs start from the closest feasible positions to their respective targets, the system achieves the highest possible sensing quality. However, this comes at the expense of communications performance. 
To balance this tradeoff, the UAVs iteratively adjust their power split factor and move toward the FBS, which is strategically positioned at the centroid of all UAVs. This adaptive approach enables the system to gradually converge to an optimal balance between sensing and communications.  
Notably, convergence is achieved within just 40 iterations, demonstrating the efficiency of our solution and its ability to reach an optimal configuration within a reasonable time frame.  Hence, this figure highlights the computational efficiency of the DJRC solution, which achieves polynomial complexity by guiding UAVs iteratively along the shortest path to the FBS based on computed rewards, avoiding exhaustive evaluations. This approach minimizes computation time, making DJRC ideal for time-sensitive post-disaster search and rescue operations.  
 \begin{figure}[t!]
	\centering
		\scalebox{1.35}{\includegraphics[width=0.3\textwidth]{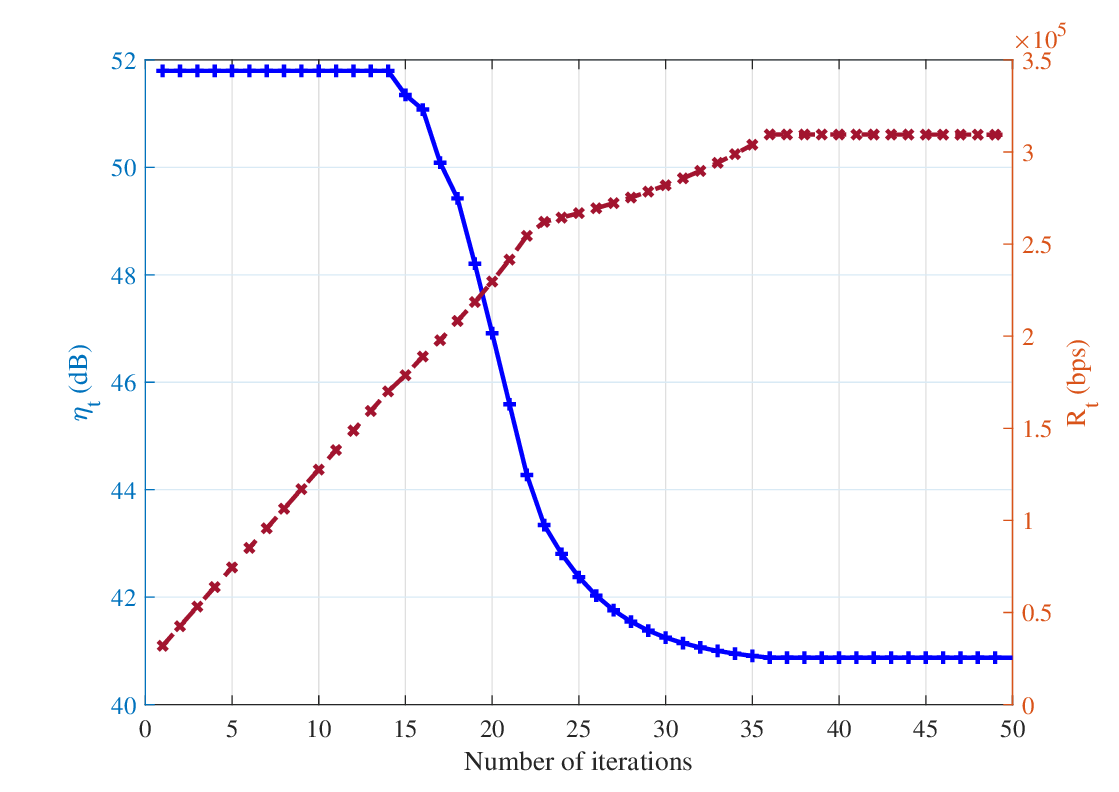}}
	\caption{ Convergence of the DJRC algorithm: evolution of total received SNR ($\eta_t$) and sum data rate (${R}_t$) over iterations.  }
	\label{fig:convergence}
\end{figure}

For comparison with the FROC and ORFC solutions, we consider two scenarios: (i) varying the number of targets, as shown in Figure \ref{fig:varyingTargets}, and (ii) varying the total available power at each UAV, as shown in Figure \ref{fig:varyingPower}.  
The first scenario assesses the performance of each method as the complexity of the target distribution increases. Figure \ref{fig:varyingTargets}  demonstrates that our DJRC solution consistently provides the best overall performance by maintaining the highest detection quality $\eta_t$ while satisfying the data rate constraint, ${R}_t$. As the number of targets increases, the resources allocated to each UAV decrease, leading to a reduction in detection quality, as shown in Figure \ref{fig:varyingTargets}-(a). Nevertheless, the DJRC algorithm continues to outperform the FROC and ORFC solutions because it jointly optimizes both radar and communications functionalities. In contrast, the FROC solution focuses solely on communications, leading to UAVs being allocated at the maximum radar range, which results in lower detection quality ($\eta_{\text{min}}$) while maximizing the sum data rate ${R}_t$ (see Figure \ref{fig:varyingTargets}-(b)). On the other hand, the ORFC solution prioritizes radar detection quality, allocating UAVs to positions that maximize radar performance while ensuring the minimum data rate constraint. However, it ignores the joint optimization of UAV locations and power split factors. 

\begin{figure}[t!]
    \centering
    \begin{subfigure}[b]{0.42\textwidth} 
        \centering
        \includegraphics[width=\textwidth]{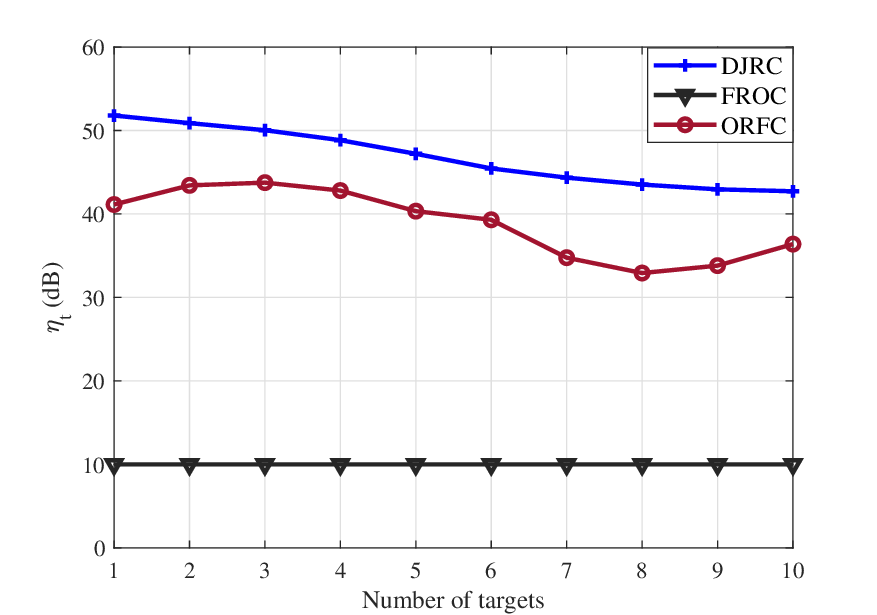}
        \caption{}
        \label{fig:subfig_a}
    \end{subfigure}
    \hfill 
    \begin{subfigure}[b]{0.42\textwidth}
        \centering
        \includegraphics[width=\textwidth]{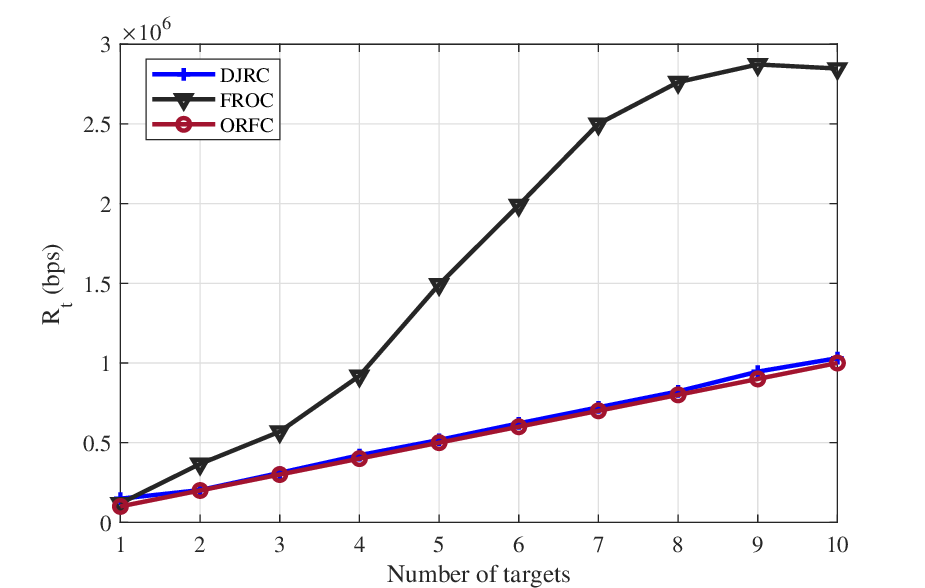}
        \caption{}
        \label{fig:subfig_b}
    \end{subfigure}
    \caption{Performance comparison of DJRC, FROC, and ORFC solutions in terms of: (a) total received SNR $\eta_t$, and (b) sum of achieved data rates ${R}_t$, under varying numbers of targets. }
    \label{fig:varyingTargets}
\end{figure}

The second scenario fixed the number of targets at 10 and varied the total available power at each UAV, $p_t$ (see Figure \ref{fig:varyingPower}). This setup assesses how the detection quality improves as the available power increases. In Figure \ref{fig:varyingPower}, the DJRC solution consistently delivers the best overall performance in terms of detection quality, $\eta_t$, while ensuring that the total data rate of all UAVs satisfies the 1 Mbit/s constraint. This demonstrates the ability of the DJRC algorithm to effectively balance both radar and communications requirements, outperforming the FROC and ORFC solutions even as power availability increases. 

\begin{figure}[t!]
    \centering
    \begin{subfigure}[b]{0.39\textwidth} 
        \centering
        \includegraphics[width=\textwidth]{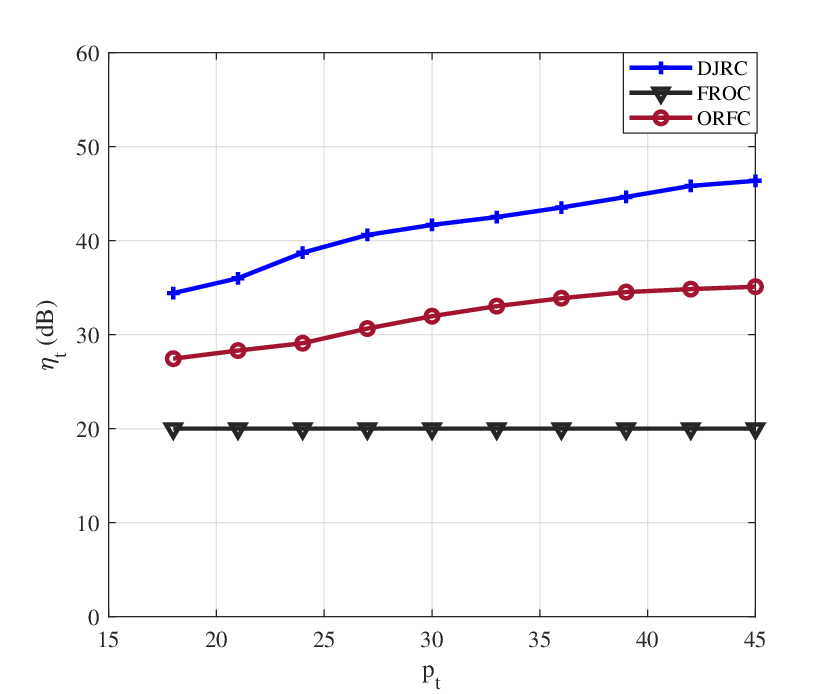}
        \caption{}
        \label{fig:subfig_a}
    \end{subfigure}
    \hfill 
    \begin{subfigure}[b]{0.4\textwidth}
        \centering
        \includegraphics[width=\textwidth]{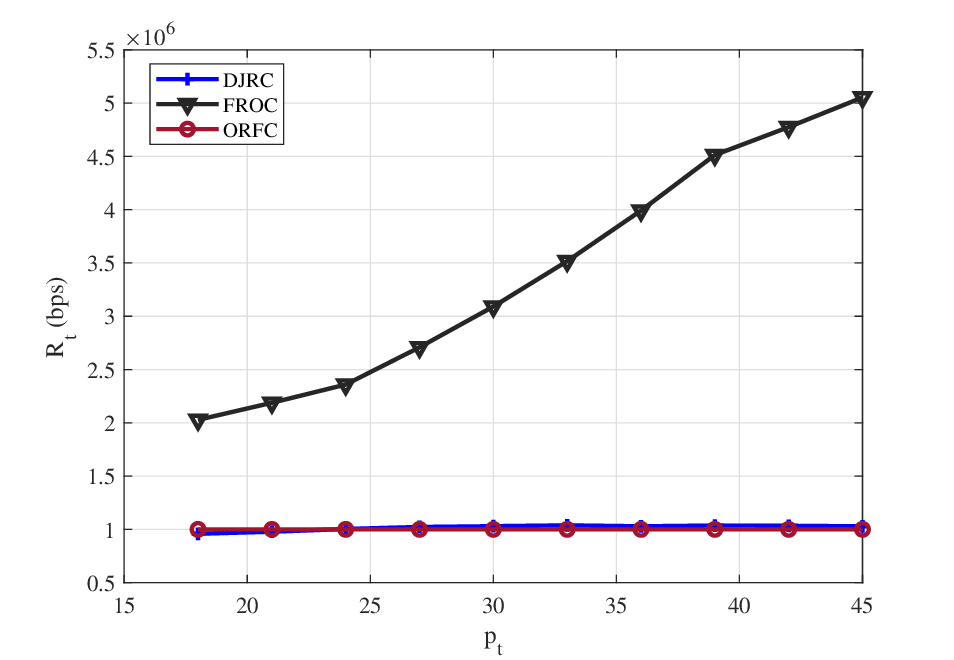}
        \caption{}
        \label{fig:subfig_b}
    \end{subfigure}
    \caption{Performance comparison of DJRC, FROC, and ORFC solutions in terms of: (a) total received SNR $\eta_t$, and (b) sum of achieved data rates ${R}_t$, under varying total power per UAV.  }
    \label{fig:varyingPower}
\end{figure}

\section{Conclusion \label{sec:conclusion}}

In this paper, we propose a Distributed Joint Radar-Communications (DJRC) solution for optimizing UAV locations in complex environments, without relying on ground-based station availability. By leveraging radar-communications UAVs, our system effectively tackles challenges commonly encountered in such environments, including damaged infrastructure and blocked roads. UAVs could enhance detection and data collection in inaccessible areas, enabling faster decisions and efficient operations, particularly in time-sensitive scenarios like disaster response. 
In particular, our approach focuses on maximizing detection quality while ensuring reliable communications quality by optimizing the power split and dynamically positioning UAVs at the best locations, striking a balance between sensing and communications. The proposed DJRC algorithm consistently delivers superior performance compared to alternative methods, while also significantly reducing computational complexity to a polynomial scale dependent on the number of UAVs, with linear dependence on the required iteration counts. This computational efficiency makes the DJRC solution highly suitable for real-time deployment in critical scenarios, where rapid decision-making is paramount. 
  
This work paves the way for future research, including integrating diverse sensor/communications technologies, optimizing UAV trajectories and resources, and developing efficient algorithms for cooperative multi-UAV sensing and communications.  
 
\balance 

\bibliographystyle{IEEEtrannames}
\bibliography{8Pages_paper}

\end{document}